\documentstyle[12pt]{article}
\parskip=\medskipamount                 
\thicklines                     
\newcommand{\bim}[6]{\bibitem{#1}#2, {\em #3\/}$\;${\bf
#4}$\;$(#5)$\;${#6}.}
\def\IR{\relax{\rm I\kern-.18em R}}
\def\ZZ{\relax{\sf Z\kern-.4em Z}}
\def\a{\alpha} \def\b{\beta}   \def\e{\epsilon} 
 \def\l{\lambda} 
   
   \def\cD{{\cal D}}
   
 \def\cK{{\cal K}} \def\cL{{\cal L}} \def\cM{{\cal M}}
 \def\cO{{\cal O}}  
 
\newtheorem{proposition}{Proposition}[section]
\newtheorem{corollary}{Corollary}[section]
\newtheorem{conjecture}{Conjecture}[section]

\newlength{\shiftwidth}
\def\shift#1{&&\hbox to \shiftwidth{\hfill $\displaystyle#1$}}
\newlength{\sshiftwidth}
\def\sshift#1{\lefteqn{\hbox to
\sshiftwidth{\hfill$\displaystyle#1$}}}
\def\llefteqn#1{\hbox to 0pt{$\displaystyle #1 $\hss}\hspace*{1in}}

\def\sign#1{{\rm sign}\left(#1\right)}

\def\emsign#1{e^{-\frac{i\pi}{4}\sign{#1}}}

\def\dim{{\rm dim}\,}
\def\Vol{{\rm Vol}\,}
\def\ord{{\rm ord}\,}

\def\Tr{{\rm Tr}}
\def\Pexp{{\rm Pexp}}
\def\hol#1{\Pexp\left(\oint_{#1} A_\mu dx^\mu\right)}
\def\thol#1#2{\Tr_{#1}\hol{#2}}

\def\Tub{{\rm Tub}}

\def\Upq{U^{(p,q)}}

\def\PU{\Phi(\Upq)}

\def\pf{\phi_{\rm fr}}
\def\Nph{N_{\rm ph}}

\def\pqn{\left(\frac{p}{q}+\nu\right)}

\def\tspqn{3\sign{\frac{p}{q}+\nu}}

\def\ztramk{Z^{({\rm tr})}_{\a}(M,\cK;k)}

\def\ztrmk{Z^{({\rm tr})}(M;k)}
\def\ztrmpk{Z^{({\rm tr})}(M^\prime;k)}

\def\rhs{RHS$\;$}
\def\rhss{RHS}

\def\ohm{\ord H_1(M,\ZZ)}

\def\va{\vec{a}}
\def\vb{\vec{b}}

\def\vgr{\vec{\rho}}
\def\vx{\vec{x}}
\def\vn{\vec{n}}
\def\vgs{\vec{\sigma}}

\def\ztraml{Z^{({\rm tr})}_{\a_1,\ldots,\a_n}(M,\cL,k)}
\def\pva{\prod_{j=1}^{n}\left(\frac{K}{4\pi}\frac{d^2\va_j}{|\va_j|}\right)}
\def\spint{\int_{|\va_j|=\frac{\a_j}{K}}\pva}
\def\adots{(\va_1,\ldots,\va_n)}
\def\lm{L_m\adots}
\def\elm{\exp\left(\frac{i\pi K}{2}\sum_{m=2}^{\infty}\lm\right)}
\def\pml{P_{m,l}\adots}
\def\spml{\sum_{\stackrel{\scriptstyle l,m=0}{l+m\neq
0}}^{\infty}K^{-m}\pml}
\def\onespml{\left[1+\spml\right]}
\def\finv{F_m(\vb_1,\ldots,\vb_m)}
\def\lt{l^{(3)}}
\def\lti{\lt_{ijk}}
\def\lf{l^{(4)}}
\def\lfi{\lf_{ij,kl}}
\def\eadots{e^{2\pi ia_1},\ldots,e^{2\pi ia_n}}
\def\mintub{M\setminus{\rm Tub}(\cL)}
\def\dal{\Delta_A(M,\cL;\eadots)}
\def\invrtl{\tau_R^{-1}(\mintub;\eadots)}
\def\mind{M_{ij,\mu\nu}}
\def\emind{e^{\frac{i\pi}{4}\sign{\mind}}}
\def\emmind{e^{-\frac{i\pi}{4}\sign{\mind}}}
\def\pl{P_{0,l}(a_1\vn,\ldots,a_n\vn)}
\def\pln{P_{0,l}(a_1\vn,\ldots,a_{n-1}\vn,0)}

\def\spl{\sum_{l=2}^{\infty}\pl}
\def\spln{\sum_{l=2}^{\infty}\pln}
\def\proda{\prod_{j=1}^{n}a_j}
\def\addots{a_1,\ldots,a_n}

\def\fpq{\frac{p}{q}}
\def\pqlf{\fpq+l_{nn}}
\def\pql{p+ql_{nn}}
\def\fpql{\frac{q}{\pql}}
\def\emsp{e^{-\frac{3}{4}i\pi\sign{\pqlf}}}
\def\skq{\sqrt{2K|q|}}
\def\fsq{\frac{2\sign{q}}{\skq}}

\def\tl{\cL_{(n,mn)}}
\def\zatl{Z_{\a_1,\ldots,\a_n}(S^3,\tl;k)}

\def\smin{S^3\setminus{\rm Tub}(\cL)}
\def\gpi{\pi_1(\smin)}
\def\lmi{l^{(\mu)}_{i_1,\ldots,i_m}}

\catcode`\@=11

\newif\if@fewtab\@fewtabtrue

\catcode`\@=11

\newif\if@fewtab\@fewtabtrue

{\count255=\time\divide\count255 by 60
\xdef\hourmin{\number\count255}
\multiply\count255 by-60\advance\count255 by\time
\xdef\hourmin{\hourmin:\ifnum\count255<10 0\fi\the\count255}}
\def\ps@draft{\let\@mkboth\@gobbletwo
    \def\@oddhead{}
    \def\@oddfoot
       {\hbox to 7 cm{$\scriptstyle Draft\ version:\ \draftdate$
       \hfil}\hskip -7cm\hfil\rm\thepage \hfil}
    \def\@evenhead{}\let\@evenfoot\@oddfoot}

\def\ceqno{\global\@fewtabfalse
    \ifcase\@eqcnt \def\@tempa{& & &}\or \def\@tempa{& &}
      \or \def\@tempa{&}
      \or\def\@tempa{}\fi\@tempa
{\rm(\theequation)}}

\def\aeqno#1{\global\@fewtabfalse
    \ifcase\@eqcnt \def\@tempa{& & &}\or \def\@tempa{& &}
      \or \def\@tempa{&}
      \or\def\@tempa{}\fi\@tempa
{\rm(\theequation,#1)}}

\def\label#1{\ifnum\draftcontrol=1
 \global\def\draftnote{$\scriptstyle #1$}\fi
 \@bsphack\if@filesw {\let\thepage\relax
   \def\protect{\noexpand\noexpand\noexpand}%
\xdef\@gtempa{\write\@auxout{\string
      \newlabel{#1}{{\@currentlabel}{\thepage}}}}}\@gtempa
   \if@nobreak \ifvmode\nobreak\fi\fi\fi
  \@esphack}

\def\alabel#1#2{\label{#1}\global\@fewtabfalse
    \ifcase\@eqcnt \def\@tempa{& & &}\or \def\@tempa{& &}
      \or \def\@tempa{&}
      \or\def\@tempa{}\fi\@tempa
{\hbox to 3cm{\phantom{\rm(\theequation,#2)}
\draftnote \hfil}\hskip -3cm {\rm(\theequation,#2)}}}

\def\clabel#1{\label{#1}\global\@fewtabfalse
    \ifcase\@eqcnt \def\@tempa{& & &}\or \def\@tempa{& &}
      \or \def\@tempa{&}
      \or\def\@tempa{}\fi\@tempa
{\hbox to 3cm{\phantom{\rm(\theequation)}
\draftnote \hfil}\hskip -3cm{\rm(\theequation)}}}

\def\eqnarray{\def\draftnote{{}}\global\@fewtabtrue
\stepcounter{equation}\let\@currentlabel=\theequation
\global\@eqnswtrue
\global\@eqcnt\z@\tabskip\@centering\let\\=\@eqncr
$$\halign to \displaywidth\bgroup\@eqnsel\hskip\@centering\@eqcnt\z@
  $\displaystyle\tabskip\z@{##}$&\global\@eqcnt\@ne
  \hskip 1\arraycolsep \hfil${##}$\hfil
  &\global\@eqcnt\tw@ \hskip 1\arraycolsep
$\displaystyle\tabskip\z@{##}$
\hfil  \tabskip\@centering&\global\@eqcnt\thr@@\llap{##}\tabskip\z@
\cr}

\def\endeqnarray{\@@eqncr\egroup
      \global\advance\c@equation\m@ne$$\global\@ignoretrue}

\def\@eqnnum{\hbox to 3cm{\phantom{\rm(\theequation)} \draftnote
                         \hfil}\hskip -3cm {\rm(\theequation)}}

\def\@@eqncr{\let\@tempa\relax
    \ifcase\@eqcnt \def\@tempa{& & &}\or \def\@tempa{& &}
      \or \def\@tempa{&}
      \or\def\@tempa{}
\fi\@tempa
\if@eqnsw
\if@fewtab\@eqnnum\fi
\stepcounter{equation}\fi\global
\@eqnswtrue\global\@eqcnt\z@\global\@fewtabtrue\cr}

\def\draftcite#1{\ifnum\draftcontrol=1#1\else{}\fi}

\def\@lbibitem[#1]#2{\item{}\hskip -3cm \hbox to 2cm
{\hfil$\scriptstyle\draftcite{#2}$}\hskip
1cm[\@biblabel{#1}]\if@filesw
     {\def\protect##1{\string ##1\space}\immediate
      \write\@auxout{\string\bibcite{#2}{#1}}}\fi\ignorespaces}

\def\@bibitem#1{\item\hskip -3cm \hbox to 2cm
{\hfil $\scriptstyle\draftcite{#1}$}\hskip 1cm
\if@filesw \immediate\write\@auxout
       {\string\bibcite{#1}{\the\value{\@listctr}}}\fi\ignorespaces}

\def\nsection#1{\section{#1}\setcounter{equation}{0}}
\def\nappendix#1{\def\thesection{A#1}\section*{Appendix #1}
\def\theequation{{A#1.\arabic{equation}}}
\def\theproposition{{A#1.\arabic{proposition}}}
\setcounter{equation}{0}
\setcounter{proposition}{0}}

\def\draftdate{\number\month/\number\day/\number\year\ \ \ \hourmin }

\global\def\draftcontrol{0}
\catcode`\@=12

\def\theequation{{\thesection.\arabic{equation}}}

\def\qq{\begin{eqnarray}}
\def\qqq{\end{eqnarray}}

\begin{document}

\begin{titlepage}
\centerline{\hfill                 UMTG-176-94}
\centerline{\hfill                 hep-th/9403021}
\vfill
\begin{center}
{\large \bf A Contribution of the Trivial Connection to the Jones
Polynomial and Witten's Invariant of 3d Manifolds II.
} \\

\bigskip
\centerline{L. Rozansky\footnote{Work supported
by the National Science Foundation
under Grant No. PHY-92 09978.
}}

\centerline{\em Physics Department, University of Miami
}
\centerline{\em P. O. Box 248046, Coral Gables, FL 33124, U.S.A.}

\vfill
{\bf Abstract}

\end{center}
\begin{quotation}

We extend the results of our previous paper~\cite{RoI} from knots to
links by using a formula for the Jones polynomial of a link derived
recently by N.~Reshetikhin. We establish a relation between the
parameters of this formula and the multivariable Alexander
polynomial. This relation is illustrated by an example of a torus
link. We check that our expression for the Alexander
polynomial satisfies some of its basic properties. Finally we derive
a link surgery formula for the loop corrections to the trivial
connection contribution to Witten's invariant of rational homology
spheres.

\end{quotation}
\vfill
\end{titlepage}

\pagebreak
\nsection{Introduction}
This paper is an expansion of our previous work~\cite{RoI}. We will
try to extend the results of that paper from knots to links. Our main
tool will be the formula for the Jones polynomial of a link proposed
recently by N.~Reshetikhin\footnote{I am indebted to N.~Reshetikhin
for communicating the results of his research.}~\cite{Re}.

We start by briefly reviewing the notations of~\cite{RoI}
(they will be used throughout this paper) as well as some of
its results. Let $\cL$ be an $n$-component link in a 3-dimensional
manifold $M$. We assign an $\a_j$-dimensional $SU(2)$ representation
to each component $\cL_j$ of $\cL$. E. Witten introduced
in~\cite{Wi1} an invariant $Z_{\a_1,\ldots,a_n}(m,\cL;k)$
which is a path integral over the gauge equivalence classes of $SU(2)$
connection $A_\mu$ on $M$:
\qq
Z_{\a_1,\ldots,\a_n}(M,\cL;k)=\int[\cD A_\mu]
\exp\left(\frac{i}{\hbar}S_{CS}\right)
\prod_{j=1}^{n}\thol{\a_j}{\cL_j},
\label{1.1}
\qqq
here $S_{CS}$ is the Chern-Simons action
\qq
S_{CS}=\frac{1}{2}\,\Tr\,\epsilon^{\mu\nu\rho}\int_{M}dx
(A_\mu \partial_\nu A_\rho + \frac{2}{3}A_\mu A_\nu A_\rho),
\label{1.2}
\qqq
$\hbar$ is a ``Planck's constant'':
\qq
\hbar=\frac{2\pi}{k},\;\;k\in \ZZ,
\label{1.3}
\qqq
the trace $\Tr$ in eq.~(\ref{1.2})
is taken in the fundamental (2-dimensional)
representation and $\thol{\a_j}{\cL_j}$ are the traces of holonomies
along the link components $\cL_j$ taken in the
$\a_j$-dimensional representations.

The path integral~(\ref{1.1}) can be calculated in the stationary
phase approximation in the limit of large $k$. The stationary points
of the Chern-Simons action~(\ref{1.2}) are flat connections and
Witten's invariant is presented as a sum over connected pieces $\cM_c$
of their moduli space $\cM$:
\begin{eqnarray}
Z_{\a_1,\ldots,\a_n}(M,\cL;k)&=&
\sum_{\cM_c}Z^{(\cM_c)}_{\a_1,\ldots,\a_n}(M,\cL;k),
\nonumber\\
Z^{(\cM_c)}_{\a_1,\ldots,\a_n}(M,\cL;k)&=&
\exp\frac{i}{\hbar}\left(S_{CS}^{(c)}+\sum_{n=1}^{\infty}
S_n^{(c)}\hbar^n\right),
\label{1.13}
\end{eqnarray}
here $S_{CS}$ is a Chern-Simons action of flat connections of $\cM_c$
and $S_n^{(c)}$ are the quantum $n$-loop
corrections to the
contribution of $\cM_c$. The 1-loop correction is a determinant of the
quadratic form describing the small fluctuations of $S_{CS}(A_\mu)$
around a stationary phase point. Its major features were determined by
Witten \cite{Wi1}, Freed and Gompf \cite{FrGo}, and Jeffrey \cite{Je}
(some further details were added in \cite{Ro1}):
\begin{eqnarray}
\lefteqn{
e^{iS_1^{(c)}}=
\frac{
(2\pi\hbar)^{\frac{\dim H_c^0-\dim H^1_c}{2}}}
{\Vol(H_c)}
\exp\left(\frac{i}{\pi} S_{CS}-\frac{i\pi}{4}\Nph
\right)
}
\label{1.14}\\
\shift{\times
\int_{\cM_c}\left[
\sqrt{|\tau_R|}
\prod_{j=1}^{n}\thol{\a_j}{\cL_j}\right],
}
\nonumber
\end{eqnarray}
here $H_c$ is an isotropy group of $\cM_c$ (i.e. a subgroup of $SU(2)$
which commutes with the holonomies of connections $A_\mu^{(c)}$ of
$\cM_c$), $\Nph$ is expressed \cite{FrGo} as
\qq
\Nph=2I_c + \dim H_c^0 + \dim H_c^1 + 3(1+b_M^1),
\label{1.014}
\qqq
$I_c$ is a spectral flow of the operator $L_-=\star D+D\star$ acting
on 1- and 3-forms, $D$ being a covariant derivative, $H^0_c$ and
$H^1_c$ are cohomologies of $D$, and $b^1_M$ is the first Betti
number of $M$. $\tau_R$ is a Reidemeister-Ray-Singer torsion. It was
observed in \cite{Je} that $\sqrt{\tau_R}$ defines a ratio of volume
forms on $\cM_c$ and $H_c$.

In a particular case of a rational homology sphere (\rhs), the 1-loop
correction to the contribution of the trivial connection is
\qq
e^{iS_1^{({\rm tr})}(M)}=\sqrt{2}\pi [K\ohm]^{-\frac{3}{2}}.
\label{3.7}
\qqq
Basing on our calculation of Witten's invariant of Seifert manifolds
we conjectured in~\cite{Ro1} that
\qq
S_2^{({\rm tr})}(M)=3\l_{CW}(M),
\label{3.8}
\qqq
here $\l_{CW}$ is Casson-Walker invariant of $M$ (it was calculated
for Seifert manifolds by C.~Lescop in~\cite{Le1}).

Witten has suggested in~\cite{Wi1} a surgery formula for the invariant
$Z(M;k)$. We need to introduce some notations in order to describe it.
We pick two basic cycles on the boundaries of the tubular
neighborhoods $\Tub(\cL_j)$ of the link components $\cL_j$. A cycle
$C_1^{(j)}$ is a meridian of $\cL_j$, it can be contracted through
$\Tub(\cL_j)$. A cycle $C_2^{(j)}$ has a unit intersection number with
$C_1^{(j)}$, it is defined only modulo $C_1^{(j)}$. We denote as
$l_{ij}$ the linking numbers of the link components. The self-linking
number $l_{jj}$ is a linking number between $\cL_j$ and $C_2^{(j)}$.

A surgery on a link component $\cL_j$ is determined by an $SL(2,\ZZ)$
matrix $U^{(p_j,q_j)}$:
\qq
U^{(p_j,q_j)}=\left(
\begin{array}{cc}
p_j&r_j\\q_j&s_j
\end{array}\right)\in SL(2,\ZZ),\;\;
p_j s_j-q_j r_j=1.
\label{1.4}
\qqq
The surgery means that we cut $\Tub(\cL_j)$ out and glue it back in
such a way that the cycles $p_j C_1^{(j)} + q_j C_2^{(j)}$ and
$r_j C_1^{(j)} + s_j C_2^{(j)}$ on the boundary of the complement
$M\setminus\Tub(\cL_j)$ are glued to the cycles $C_1^{(j)}$ and
$C_2^{(j)}$ on the boundary of $\Tub(\cL_j)$.

Let $M^\prime$ be a manifold constructed by $n$ surgeries
$U^{(p_j,q_j)}$ on the components of the link $\cL$. Then, according
to~\cite{Wi1},
\qq
Z(M^\prime;k)=e^{i\pf}
\sum_{\a_1\ldots,\a_n=1}^{k+1}
Z_{\a_1\ldots,\a_n}(M,\cL;k)
\prod_{j=1}^n\tilde{U}_{\a_j 1}^{(p_j,q_j)},
\label{1.12}
\qqq
here $\tilde{U}_{\a\b}^{(p,q)}$ is a representation of the group
$SL(2,\ZZ)$ in the $k+1$-dimensional space of affine $SU(2)$
characters:
\begin{eqnarray}
&{\displaystyle
\tilde{U}^{(p,q)}_{\a\b}=i\frac{\sign{q}}{\sqrt{2K|q|}}
e^{-\frac{i\pi}{4}\PU}
\sum_{\mu=\pm 1}\sum_{n=0}^{q-1}
\mu\exp\frac{i\pi}{2Kq}\left[p\a^2-2\a(2Kn+\mu\b)
+s(2Kn+\mu\b)^2\right],
}
&
\nonumber\\
&1\leq\a,\b\leq K-1, \;\;\;\;\;\;\;K=k+2&
\label{1.9}
\end{eqnarray}
(see e.g.~\cite{Je} and references therein),
$\PU$ is the Rademacher function:
\qq
\Phi\left[
\begin{array}{cc}p&r\\q&s\end{array}
\right]=\frac{p+s}{q}-12s(p,q),
\label{1.7}
\qqq
$s(p,q)$ is a Dedekind sum:
\qq
s(p,q)=\frac{1}{4q}\sum_{j=1}^{n-1}
\cot\left(\frac{\pi j}{q}\right)
\cot\left(\frac{\pi pj}{q}\right).
\label{1.8}
\qqq
$\pf$ is a framing correction (all Witten's invariants are reduced
to the canonical framing, see e.g.~\cite{FrGo}):
\qq
\pf=\frac{\pi}{4}\frac{K-2}{K}\left[
\sum_{j=1}^n\Phi(U^{(p_j,q_j)})-3\sign{L^{({\rm tot})}}
\right],
\label{1.11}
\qqq
here $L^{({\rm tot})}$ is an $n\times n$ matrix
\qq
L^{({\rm tot})}_{ij}=l_{ij}+\frac{p_j}{q_j}\delta_{ij}.
\label{my.1}
\qqq
The mathematical proof of the invariance of eq.~(\ref{1.12}) was given
by N.~Reshetikhin and V.~Turaev~\cite{ReTu}. They also formulated
general conditions on the elements of that formula that would
guarantee its invariance.

In our previous paper~\cite{RoI} we gave a ``path-integral'' proof of
the following conjecture which P.~Melvin and H.~Morton~\cite{MeMo}
formulated for $M=S^3$:
\begin{proposition}
\label{pmy.1}
The trivial connection contribution to the Jones polynomial of a knot
$\cK$ in a \rhs $M$ can be expressed as
\qq
\ztramk =\ztrmk\,\exp\left[\frac{i\pi}{2K}\nu(\a^2-1)\right]
\,\a J(\a,K),
\label{2.15}
\qqq
here $\nu$ is a self-linking number of $\cK$ and $J(\a,K)$ is a
function that has the following expansion in $K^{-1}$ series:
\qq
J(\a,K)=\sum_{n=0}^{\infty}\sum_{m=0}^{n}
D_{m,n}\a^m K^{-n}.
\label{2.16}
\qqq
The dominant part of this expansion is related to the Alexander
polynomial of $\cK$:
\qq
\pi a\sum_{n=0}^{\infty}D_{n,n}a^n=
[\ohm]\frac{\sin\left(\frac{\pi a}{m_2 d}\right)}
{\Delta_A \left(M,\cK;e^{2\pi i\frac{a}{m_2 d}}\right)},
\label{my.2}
\qqq
the integer numbers $m_2$ and $d$ are defined in~\cite{RoI}, $m_2=d=1$
if $M=S^3$.
\end{proposition}

We combined the results of this proposition with the finite Poisson
resummation formula in order to derive a knot surgery formula for the
loop corrections to the trivial connection contribution to Witten's
invariant of a \rhs:
%
\begin{proposition}
If $M$ and $M^\prime$ are rational homology spheres and $M^\prime$ is
constructed by a rational surgery $\Upq$ on a knot $\cK$ in $M$, which
has a self-linking number $\nu$, then the trivial connection
contribution to Witten's invariants of $M$ and $M^\prime$ are related
by the formula
\begin{eqnarray}
\lefteqn{\ztrmpk=}
\label{3.*4}\\
&&=\ztrmk
\frac{\sign{q}}{\sqrt{2K|q|}}
e^{-i\frac{3}{4}\pi\sign{\frac{p}{q}+\nu}}
\exp\left[\frac{i\pi}{2K}\left(12s(p,q)-\pqn+\tspqn\right)\right]
\nonumber\\
\shift{
\times\int^{+\infty}_{\stackrel{-\infty}{[\a_*=0]}}
d\a\,\sin\left(\frac{\pi\a}{Kq}\right)\,\a\,J(\a,K)\,
\exp\left[\frac{i\pi}{2K}\pqn\a^2\right],
}
\nonumber
\end{eqnarray}
here the function $J(\a,K)$ comes from eq.~(\ref{2.15}), it is a
Feynman diagram contribution of the trivial connection to the Jones
polynomial of $\cK$.
The integral
$\int^{+\infty}_{\stackrel{-\infty}{[\a_*=0]}}$ in eq.~(\ref{3.*4})
should be calculated in the following way: the preexponential factor
$\sin\left(\frac{\pi\a}{Kq}\right)\,\a\,J(\a,K)$ should be expanded in
$K^{-1}$ series with the help of eq.~(\ref{2.16}), then each term
should be integrated separately with the gaussian factor
$\exp\left[\frac{i\pi}{2K}\pqn\right]$.
\label{p3.1}
\end{proposition}
%
%
\begin{corollary}
Only a finite number of Vassiliev's invariants participate in a
surgery formula for $\ztrmpk$ at a given loop order.
\label{c3.1}
\end{corollary}
%
A 2-loop part of eq.~(\ref{3.*4}) coincides with Walker's surgery
formula for Casson-Walker invariant. This proves the conjectured
relation~(\ref{3.8}).

A generalization of eq.~(\ref{2.15}) for links was
derived recently by N.~Reshetikhin\footnote{I am indebted to
N.~Reshetikhin for sharing the results of his unpublished
research.}~\cite{Re}. He observed that if the dimensions $\a_i$ in
eq.~(\ref{1.1}) are big enough, then the representation spaces can be
treated classically: the matrix elements of Lie algebra generators in
$\a_j$-dimensional representation can be substituted by functions on
the coadjoint orbit of radius $\a_j$ and a trace over the
representation can be substituted by an integral over that orbit.
\begin{proposition}
\label{pf2.1}
Let $\cL$ be an $n$-component link in a \rhs $M$. Then the trivial
connection contribution to its Jones polynomial can be expressed as a
multiple integral over the $SU(2)$ coadjoint orbits:
\begin{eqnarray}
\lefteqn{\ztraml=\ztrmk\spint\elm}
\label{5.1}\\
\shift{
\times\onespml.}
\nonumber
\end{eqnarray}
here $\va_j$ are 3-dimensional vectors with fixed length
\qq
|\va_j|=\frac{\a_j}{K}
\label{5.2}
\qqq
and $\lm$, $P_{l,m}\adots$ are homogeneous invariant (under $SO(3)$
rotations) polynomials of degree $M$. In particular,
\qq
L_2\adots=\sum_{i,j=1}^{n}l_{ij}\,\va_i\cdot\va_j,
\label{5.4}
\qqq
$l_{ij}$ is the linking number of the link components $\cL_i$ and
$\cL_j$.
\end{proposition}
An example of this formula for a torus link is derived in Appendix~1.

In our paper~\cite{Ro2} we proved this proposition by deriving a set
of Feynman rules to calculate the coefficients of the polynomials
$L_m$ and $P_{ml}$. These rules allowed us to establish the following
property of the polynomials $L_m$
\begin{proposition}
\label{P5.1}
The polynomials $\lm$ are produced from invariant homogeneous
polynomials $\finv$ of order $m$ by substituting $n$ vectors $\va_j$
in place of $m$ vectors $\vb_j$. The polynomials $\finv$, $m\geq 3$
are equal to zero if at least $m-1$ of $m$ vectors $\vb_j$ are
parallel.
\end{proposition}
We also conjectured a relation between the coefficients of the
polynomials $L_m$ and Milnor's linking numbers:
\begin{conjecture}
\label{cf3.1}
If $L_l\adots=0$ for all $l<m$, then the coefficients of the
polynomial $\lm$ are proportional to the $m$th order Milnor's linking
numbers $\lmi$ of the link $\cL$:
\qq
\lm=\frac{(i\pi)^{m-2}}{m}
\sum_{1\leq i_1,\ldots,i_m\leq n}
\lmi\Tr (\vgs\cdot\va_{i_i})
\cdots(\vgs\cdot\va_{i_m}),
\label{n.1}
\qqq
here $\vgs=(\sigma_1,\sigma_2,\sigma_3)$ is a 3-dimensional vector
formed by Pauli matrices. \end{conjecture}

We will need especially the polynomials $L_3$, $L_4$
and $P_{0,2}$:
\begin{eqnarray}
L_3\adots&=&
\sum_{i,j,k=1}^{n}\lti\;\va_i\cdot(\va_j\times\va_k),
\label{5.5}\\
L_4\adots&=&\sum_{i,j,k,l=1}^{n}\lfi\;
(\va_i\times\va_j)\cdot(\va_k\times\va_l).
\label{5.6}\\
P_{0,2}\adots&=&\sum_{i,j=1}^{n}p_{ij}\va_i\cdot\va_j.
\label{5.3}
\end{eqnarray}
We demonstrated in~\cite{Ro2} that the coefficients $\lti$ and $\lfi$
are proportional to triple and quartic Milnor's linking numbers.

An obvious condition
\qq
Z^{({\rm tr})}_{1,\ldots,1}(M,\cL;k)=\ztrmk
\label{5.7}
\qqq
imposes a relation between the polynomials $L_m$ and $P_{m,l}$. It
allows us
to express the numbers $P_{m,0}$ through the coefficients of
other polynomials . For example,
\qq
P_{1,0}=-\frac{i\pi}{2}\sum_{j=1}^{n}l_{jj}.
\label{5.8}
\qqq

In this paper we will extend the Propositions~\ref{pmy.1}
and~\ref{p3.1} to links by using
the Reshetikhin's formula~(\ref{5.1}) for the Jones
polynomial of a link as a generalization of eq.~(\ref{2.15}).
In Section~\ref{s4} we derive a formula for
the multivariable Alexander polynomial of a link
in terms of the components of
Reshetikhin's formula~(\ref{5.1}) (Proposition~\ref{P5.2}). In
Section~\ref{s5} we calculate the first terms in Taylor series
expansion of the multivariable Alexander polynomial. In
Section~\ref{s6} we check that the Alexander polynomial as given by
eq.~(\ref{5.22}) does satisfy some of its basic properties
(Propositions~\ref{P5.03} and~\ref{P5.5}). In Section~\ref{s7}
we use Reshetikhin's formula in order to derive the link surgery
formula for loop corrections to the trivial connection contribution to
Witten's invariant of a \rhs (Proposition~\ref{P5.6}).
In Appendix~1
we derive Reshetikhin's presentation for the Jones polynomial of a
torus link. In Appendix~2 we give a brief description of the
structure of the moduli space of flat connections in a link complement
in the vicinity of the trivial connection. We demonstrate that those
connections are in one-to-one correspondence with the stationary
points of the phase in Reshetikhin's formula
(Proposition~\ref{pA3.1}) at least in the linear approximation around
the trivial connection.

\nsection{The Multivariable Alexander Polynomial}
\label{s4}

We will follow the method of Section~2
of~\cite{RoI} in order to relate
eq.~(\ref{5.1}) to the multivariable Alexander polynomial which we
define here as the inverse of the Reidemeister-Ray-Singer torsion of
the link complement:
\qq
\dal=\invrtl,
\label{5.9}
\qqq
here $e^{2\pi ia_i}$ are the holonomies
of the $U(1)$ flat connection
in $\mintub$ around the meridians $C_1^{(j)}$ of the link components
$\cL_j$.

We take the limit $K\longrightarrow\infty$ of the integral in
eq.~(\ref{5.1}) while keeping the ratios $\a_i/K$ fixed. Then
according to eq.~(\ref{1.13}) the partition function can be presented
as a sum over flat connections in the link complement which satisfy
(up to a conjugation) the boundary condition for each
meridian $C_1^{(j)}$:
\qq
\Pexp\,\left(\oint_{C_1^{(j)}}A_{\mu}dx^{\mu}\right)
=\exp\left(\frac{2\pi i}{K}\a_j\right),\;\;\;\;
1\leq j\leq n.
\label{5.10}
\qqq
In contrast to the knot complement considered in Section~2
of~\cite{RoI},
there may be irreducible flat connections in $\mintub$ even if the
phases $\a_i/K$ are arbitrarily small (see, e.g. Appendix~2).
Besides, there is not just one but $2^{n-1}$ reducible flat
connections due to the fact that a diagonal $SU(2)$ holonomy fixed up
to a conjugation by eq.~(\ref{5.10}) corresponds to two $U(1)$
holonomies related by a Weyl reflection, i.e. differing by the sign of
the exponent (the overall change of signs however does not change the
gauge equivalence class of the $SU(2)$ connection).

We calculate the integral of eq.~(\ref{5.1}) by the stationary phase
approximation method. Let us first assume that all $|\va_j|\ll 1$.
Then we should look for the extrema of the quadratic form~(\ref{5.4})
constrained by conditions~(\ref{5.2}). These extrema satisfy equations
\qq
\left(\sum_{j=1}^{n}l_{ij}\va_j\right)\times \va_i=0,\;\;\;\;
1\leq i\leq n.
\label{5.11}
\qqq
The solutions to these equations do indeed correspond (up to an
overall $SO(3)$ rotation) to flat connections in the link complement
for small phases $|\va_j|$ (see Appendix~1, for more details on flat
connections in the link complement see~\cite{Ro2}).
Equations~(\ref{5.11}) are obviously satisfied when all the vectors
$\va_j$ are parallel:
\qq
\va_j^{(0)}=a_j \vn,
\label{5.12}
\qqq
here $\vn$ is a unit vector and
\qq
|a_j|=\frac{\a_j}{K}.
\label{5.012}
\qqq
There are $2^{n-1}$ such inequivalent configurations depending on the
choice of signs for $a_j$ in eqs.~(\ref{5.012}). They correspond to
$2^{n-1}$ reducible flat connections. If the phases $|\va_j|$ are not
small, then we should also account for the higher order polynomials
$L_m,\;\;m\geq 3$. However Conjecture~\ref{P5.1} guarantees that the
parallel configurations~(\ref{5.12}) still remain
the stationary phase
points of the full phase in eq.~(\ref{5.1}).

The arguments of Section~2
of~\cite{RoI} suggest that the 1-loop (that is,
leading in the $K^{-1}$ expansion)
approximation to the contribution of
``reducible'' stationary phase point~(\ref{5.12}) to the
integral~(\ref{5.1}) is proportional to the Reidemeister-Ray-Singer
torsion of the link complement and inversely proportional to the
multivariable Alexander polynomial as defined by eq.~(\ref{5.9}). To
obtain this approximation we introduce the local coordinates
$\vx_j$ in the vicinity of the stationary phase point~(\ref{5.12}):
\qq
\va_j=\va_j^{(0)}+a_j\vx_j+
\frac{1}{2}\va_j^{(0)}\vx_j^2+\cO(x^3),
\;\;\;\;\vn\cdot\vx_j=0.
\label{5.13}
\qqq
We may retain only a quadratic part of the exponent in
eq.~(\ref{5.1}):
\qq
\frac{i\pi K}{2}
\sum_{i,j=1}^{n}\sum_{\mu,\nu=1}^{2}\mind\adots
x_\mu^{(i)}x_\nu^{(j)},
\label{5.14}
\qqq
here $x_\mu^{(j)}$ are coordinates of the vectors $\vx_j$. A quadratic
form $\mind$ may receive contributions from all the polynomials $L_m$:
\begin{eqnarray}
\llefteqn{\sum_{\mu,\nu=1}^{2}\mind\adots
x_\mu^{(i)}x_\nu^{(j)}=}
\label{5.15}
\\
&&
=L_{ij}\,\vx_i\cdot\vx_j+
\sum_{m=3}^{\infty}
L_m(a_1\vn,\ldots,a_i\vx_i,\ldots,a_j\vx_j,\ldots,a_n\vn).
\nonumber
\end{eqnarray}
The matrix $L_{ij}$ comes from $L_2$:
\qq
L_{ij}=l_{ij}a_i a_j-\delta_{ij}
\sum_{\stackrel{\scriptstyle k=1}{k\neq i}}^{n}
l_{ik}a_i a_k.
\label{5.16}
\qqq
In our approximation the integration measure for $\vx_i$ is reduced to
\qq
\prod_{j=1}^{n}\frac{K}{4\pi}|a_j|d^2\vx_j.
\label{5.17}
\qqq
Also we should retain only the following part of the preexponential
factor in eq.~(\ref{5.1}):
\qq
1+\spl
\label{5.18}
\qqq
(the polynomials $\pl$ do not depend on the
orientation of $\vn$).  What remains is a gaussian integral
over $\vx_j$, which would produce a square root of the determinant of
the $2n\times 2n$ matrix $\mind$. However there is a small problem:
this matrix has two zero modes:
\qq
x_{\mu}^{(j)}=\delta_{\mu 1},\;\;1\leq j\leq n\;\;\;\;\;\;
{\rm and}\;\;\;\;\;\;x_{\mu}^{(j)}=\delta_{\mu 2},\;\;1\leq j\leq n,
\label{5.19}
\qqq
which originate from $SO(3)$ rotations of $\vn$.
Zero modes appear quite often in calculations of the Alexander
polynomial.
They should
be removed from the determinant and the integration over the direction
of $\vn$ should be performed with an appropriate measure. The removal
of the zero modes is achieved either by taking a second derivative of
the characteristic polynomial of $\mind$ at zero, or by taking any of
non-zero second rank minors of two diagonal elements:
\qq
{\det}^{\prime}\mind=\left.\frac{1}{2}\partial_x^2
\det(\mind+x\delta_{ij}\delta_{\mu\nu})\right|_{x=0}
\equiv n^2\det M^{\prime\prime},
\label{5.20}
\qqq
here $M^{\prime\prime}$ is a $(2n-2)\times(2n-2)$ matrix obtained from
$\mind$ by crossing out the columns and rows to which the two diagonal
elements $M_{ii,11}$ and $M_{jj,22}$ belong ($\det M^{\prime\prime}$
does not depend on the choice of $i$ and $j$). Finally after using
eq.~(\ref{3.7}) as the 1-loop formula for $\ztrmk$ we get the
following formula for the contribution to
the Jones polynomial coming from
the reducible flat connection related to the
configuration~(\ref{5.12}):
\begin{eqnarray}
\lefteqn{
Z^{({\rm red})}_{(a_1,\ldots, a_n)}(M,\cL;k)=
\exp\left(\frac{i\pi K}{2}\sum_{i,j=1}^{n}l_{ij}a_i a_j\right)
\frac{1}{\sqrt{2K}}\emind[\ohm]^{-\frac{3}{2}}
}
\label{5.21}\\
\shift{
\times(2\pi)^{2-n}\left(\prod_{i=1}^{n}|a_i|\right)
|\det M^{\prime\prime}|^{-\frac{1}{2}}\left(1+\spl\right),
}\nonumber
\end{eqnarray}
here $\sign{\mind}$ is the difference between the numbers of positive
and negative eigenvalues of $\mind$. It is easy to relate the factors
of this expression to those of eq.~(\ref{1.14}):
$\pi^2\sum_{i,j=1}^{n}l_{ij}a_i a_j$ is the classical Chern-Simons
action, $1/\sqrt{2K}$ is the factor $\frac{\sqrt{2\pi\hbar}}{\Vol
U(1)}$ and $[\ohm]^{-1/2}$ is the contribution of the diagonal part of
$SU(2)$ to the square root of the Reidemeister-Ray-Singer torsion.
What remains is (up to a phase) the $U(1)$ torsion. According to
eq.~(\ref{5.9}) its inverse is the multivariable Alexander polynomial.
%
\begin{proposition}
The formula for the multivariable Alexander polynomial of the link
$\cL$ is
\begin{eqnarray}
\lefteqn{
\dal=-i\emmind[\ohm](2\pi)^{n-2}\frac{|\det
M^{\prime\prime}|^{\frac{1}{2}}}
{\proda}
}\label{5.22}\\
\shift{
\times\left[1+\spl\right]^{-1},
}\nonumber
\end{eqnarray}
here $\det M^{\prime\prime}$ may be expressed through the
characteristic polynomial of the matrix $\mind$ according to
eq.~(\ref{5.20}), while the matrix $\mind$ itself is expressed by
eq.~(\ref{5.15}).
\label{P5.2}
\end{proposition}
%

\nsection{Taylor Series}
\label{s5}

The bilinear form $\sum_{i,j=1}^{n}\sum_{\mu\nu=1}^{2}\mind
x_\mu^{(i)}x_\nu^{(j)}$ includes only two basic bilinear structures
coming from the r.h.s. of eq.~(\ref{5.15}): $\vx_i\cdot\vx_j$ and
$\vn\cdot(\vx_i\times\vx_j)$.
Therefore the matrix $M_{ij,\mu\nu}$ has the following block
structure:
\qq
M=
\left(
\begin{array}{c|c}
A&B\\
\hline
-B&A
\end{array}
\right),
\qqq
here $A$ and $B$ are a symmetric and an antisymmetric $n\times n$
matrices. As a result,
the characteristic polynomial
of $\mind$ is a square of another polynomial of $x$ and matrix
elements $\mind$. Also a matrix element $\mind$ is proportional to
$a_i$ and $a_j$. This together with the particular form of the zero
modes~(\ref{5.19}) guarantees that $\det M^{\prime\prime}$ is
proportional to $\left(\proda\right)^2$. Thus we conclude that the
r.h.s. of eq.~(\ref{5.22}) can be expanded in Taylor series in phases
$a_i$. The first two terms of this expansion are used in C.~Lescop's
surgery formula for Casson-Walker invariant, so we are going to find
their expression.

To get the first term in Taylor series we retain only the terms
$L_{i,j}\vx_i\cdot\vx_j$ in the r.h.s. of eq.~(\ref{5.15}). Then the
matrix $\mind$ splits into a direct sum of two equal matrices
$L_{ij}$:
\qq
\mind=L_{ij}\delta_{\mu\nu}+\cO(a^3).
\label{5.23}
\qqq
Therefore
\qq
i\emmind|\det M^{\prime\prime}|^{\frac{1}{2}}
=\frac{i^{n-2}}{n}{\det}^{\prime}L=\left.\frac{i^{n-2}}{n}
\partial_x\det(L_{ij}+x\delta_{ij})\right|_{x=0}
=i^{n-2}\det L^{\prime},
\label{5.24}
\qqq
here $L^{\prime}$ is any of the minors of diagonal elements of
$L_{ij}$ (they are all equal). Thus the first term in Taylor
expansion of the Alexander polynomial is a polynomial in $a_i$ of
degree $n-2$:
\qq
\Delta_A^{(n-2)}(M,\cL;\addots)=
-(-2\pi i)^{n-2}[\ohm]\frac{\det L^{\prime}(\addots)}
{\proda}.
\label{5.25}
\qqq
This expression coincides\footnote{I am thankful to C.~Lescop for
checking this.} with the formula of \cite{Le2}.

Obviously, eq.~(\ref{5.25}) provides the leading term in the Taylor
series expansion of the multivariable Alexander polynomial if $\det
L^{\prime}(\addots)$ is non-zero. If the linking numbers $l_{ij}$
are zero then, in view of Conjecture~\ref{cf3.1}, the dominant term
will be expressed through higher Milnor's invariants.

To get the second term in Taylor expansion we have to account for the
polynomials $L_3$ and $L_4$ in the r.h.s. of eq.~(\ref{5.15}) as well
as for the polynomial $P_{0,2}$ in the preexponential factor of
eq.~(\ref{5.21}). We can expand the exponential of eq.~(\ref{5.1}) in
$L_3$ and $L_4$. Only the second power of $L_3$ and the first power of
$L_4$ contribute to the leading power in $K$. We make a simple
rearrangement
\begin{eqnarray}
(\va_{i_1}\times\va_{i_2})\cdot
(\va_{j_1}\times\va_{j_2})&=&\det\left(
\begin{array}{cc}
(\va_{i_1}\cdot\va_{j_1})&(\va_{i_1}\cdot\va_{j_2})\\
(\va_{i_2}\cdot\va_{j_1})&(\va_{i_2}\cdot\va_{j_2})
\end{array}\right),
\label{5.26}
\qqq
\qq
[\va_{i_1}\cdot(\va_{i_2}\times\va_{i_3})]\,
[\va_{j_1}\cdot(\va_{j_2}\times\va_{j_3})]&=&
\det\left(\begin{array}{ccc}
(\va_{i_1}\cdot\va_{j_1})&(\va_{i_1}\cdot\va_{j_2})&(\va_{i_1}\cdot\va_{j_3})\\
(\va_{i_2}\cdot\va_{j_1})&(\va_{i_2}\cdot\va_{j_2})&(\va_{i_2}\cdot\va_{j_3})\\
(\va_{i_3}\cdot\va_{j_1})&(\va_{i_3}\cdot\va_{j_2})&(\va_{i_3}\cdot\va_{j_3})
\end{array}\right).
\label{5.026}
\end{eqnarray}
Multiplying the preexponential factor of the integral of
eq.~(\ref{5.1}) by an extra scalar product $(\va_i\cdot\va_j)$
is equivalent to taking a
derivative $\partial_{l_{ij}}$. By applying this trick to the
factors~(\ref{5.26}) and ~(\ref{5.026}) we arrive at the formula
\begin{eqnarray}
\lefteqn{
\Delta_A^{(n)}(M,\cL;\addots)=
-(-2\pi i)^{n-2}[\ohm]
\left[-\sum_{i,j=1}^{n}p_{i,j}a_i a_j\right.
}
\label{5.27}
\\
\shift{\left.
+
4\sum_{i_1,i_2,j_1,j_2=1}^{n}\lf_{i_1 i_2,j_1 j_2}a_{i_1}a_{j_1}
\partial_{l_{i_2 j_2}}-\frac{9}{2}\sum_{i_1,i_2,i_3,j_1,j_2,j_3=1}^{n}
\lt_{i_1 i_2 i_3}
\lt_{j_1 j_2 j_3}a_{i_1}a_{j_1}\partial_{l_{i_2 j_2}}
\partial_{l_{i_3 j_3}}\right]\frac{\det L^{\prime}}{\proda}.
}
\nonumber
\end{eqnarray}
We consider $l_{ij}$ and $l_{ji}$ as independent variables when we
take derivatives in this formula. The coefficients $p_{ij}$, $\lt_{i_1
i_2 i_3}$ and $\lf_{i_1 i_2,j_1 j_2}$ come from
eqs.~(\ref{5.3}),(\ref{5.5}) and~(\ref{5.6}).
%
\begin{proposition}
The expression~(\ref{5.22}) for the multivariable Alexander polynomial
can be expanded in Taylor series in phases $a_j$:
\qq
\dal=\sum_{j=0}^{\infty}\Delta_A^{(n-2+2j)}
(M,\cL;\addots).
\label{5.28}
\qqq
Each term $\Delta_A^{(n-2+2j)}(M,\cL;\addots)$ is a polynomial of
degree $n-2+2j$. The first two terms in this expansion are given by
eqs.~(\ref{5.25}) and~(\ref{5.27}).
\label{P5.3}
\end{proposition}
%

\nsection{Basic properties of the Alexander Polynomial}
\label{s6}
We are going to check whether the r.h.s. of eq.~(\ref{5.1}) satisfies
some basic properties of the multivariable Alexander polynomial. Let
us find the value of $\dal$ when $a_n=0$. Consider the matrix
$\mind^{\prime\prime}$. Suppose for simplicity that the diagonal
elements $M_{nn,11}$ and $M_{nn,22}$ do not belong to the two columns
and rows that were removed from $\mind$. Then it is not hard to see
that the part of $\det M^{\prime\prime}$ which is proportional only to
the second power of $a_n$, must include both these elements. As a
result,
\begin{eqnarray}
\lefteqn{
\Delta_A(M,\cL;e^{2\pi ia_1},\ldots,e^{2\pi ia_{n-1}},1)
=-i\emsign{M_{[n]}}(2\pi)^{n-2}[\ohm]}
\label{5.29}\\
\shift{
\times\left(1+\spl\right)^{-1}\frac{
|\det M^{\prime\prime}_{[n]}(a_1,\ldots,a_{n-1})|^{\frac{1}{2}}}
{\prod_{j=1}^{n-1}a_j}
\left(i\sum_{j=1}^{n-1}l_{jn}a_j\right),
}
\nonumber
\end{eqnarray}
here $M^{\prime\prime}_{[n]}$ is a $(n-4)\times(n-4)$ matrix obtained
by ``reducing'' the $(n-2)\times(n-2)$ matrix $M^{\prime\prime}$: two
rows and two columns containing the elements $M_{nn,11}$ and
$M_{nn,22}$ are removed and
$a_n=0$ is substituted in all other matrix
elements.

Suppose now that we remove the $n$th component $\cL_n$ of the link
$\cL$. We denote the remaining link as $\cL_{[n]}$. To calculate its
Jones polynomial we have to substitute $\a_n=1$ in eq.~(\ref{5.1}).
Then $|\va_n|=1/K$ and the contribution of the
configuration~(\ref{5.12}) for $1\leq j\leq n-1$ to the
integral~(\ref{5.1}) in the leading order in $K$ is equal to

\begin{eqnarray}
\lefteqn{
Z^{({\rm red})}_{(a_1,\ldots, a_{n-1})}(M,\cL_{[n]};k)=
}
\label{5.030}\\
&&=
\exp\left(\frac{i\pi K}{2}\sum_{i,j=1}^{n-1}l_{ij}a_i a_j\right)
\frac{1}{\sqrt{2K}}e^{\frac{i\pi}{4}\sign{M_{[n]}}}
[\ohm]^{-\frac{3}{2}}
(2\pi)^{3-n}
\frac{\prod_{j=1}^{n-1}|a_j|}
{|\det M^{\prime\prime}_{[n]}|^{\frac{1}{2}}}
\nonumber\\
\shift{
\times
\left(1+\spln\right)
\int_{|\vgr=1|}\frac{d^2\vgr}{4\pi}
\exp\left[i\pi\left(\sum_{j=1}^{n-1}l_{jn}a_j\right)\vgr\cdot\vn\right]
}
\nonumber\\
&&=
\exp\left(\frac{i\pi K}{2}\sum_{i,j=1}^{n-1}l_{ij}a_i a_j\right)
\frac{1}{\sqrt{2K}}e^{\frac{i\pi}{4}\sign{M_{[n]}}}
[\ohm]^{-\frac{3}{2}}
(2\pi)^{3-n}
\frac{\prod_{j=1}^{n-1}|a_j|}
{|\det M^{\prime\prime}_{[n]}|^{\frac{1}{2}}}
\nonumber\\
\shift{
\times
\left(1+\spln\right)
\frac{
\sin\left(\pi\sum_{j=1}^{n-1}l_{jn}a_j\right)}
{\pi\sum_{j=1}^{n-1}l_{jn}a_j}.
}
\nonumber
\end{eqnarray}
After extracting the $U(1)$ Reidemeister-Ray-Singer torsion from this
expression we find that
\begin{eqnarray}
\lefteqn{
\Delta_A(M,\cL_{[n]};e^{2\pi ia_1},\ldots,\e^{2\pi ia_{n-1}})=
}
\label{5.30}\\
&&=-i\emsign{M_{[n]}}(2\pi)^{n-3}[\ohm]
\left(1+\spln\right)^{-1}
\nonumber\\
\shift{
\times\frac{|\det M^{\prime\prime}_{[n]}|^{\frac{1}{2}}}
{\prod_{j=1}^{n-1}a_j}
\frac{\pi\sum_{j=1}^{n-1}l_{jn}a_j}
{\sin\left(\pi\sum_{j=1}^{n-1}l_{jn}a_j\right)}.
}
\nonumber
\end{eqnarray}
Comparing eqs.~(\ref{5.29}) and~(\ref{5.30}) we conclude that
%
\begin{proposition}
The multicolored Alexander polynomial as defined by eq.~(\ref{5.22})
satisfies the following property:
\qq
\Delta_A(M,\cL;e^{2\pi ia_1},\ldots,e^{2\pi ia_{n-1}},1)=
2i\sin\left(\pi\sum_{j=1}^{n-1}l_{jn}a_j\right)
\Delta_A(M,\cL_{[n]};e^{2\pi ia_1},\ldots,e^{2\pi ia_{n-1}}),
\label{5.31}
\qqq
here $\cL_{[n]}$ is the link $\cL$ with $n$th component removed.
\label{P5.03}
\end{proposition}
%

Now let us see what happens if we perform a $\Upq$ surgery on the
$n$th component of $\cL$ thus constructing a new \rhs $M^{\prime}$
with the link $\cL_{[n]}$ inside it. According to the surgery
formula~(\ref{1.12}) and the results of Section~3
of~\cite{RoI},
\begin{eqnarray}
\lefteqn{
Z^{({\rm tr})}_{\a_1,\ldots,\a_{n-1}}
(M^{\prime},\cL_{[n]};k)=
}
\label{5.32}\\
&&=
\ztrmk\emsp
\fsq
\exp\frac{i\pi}{2K}\left[12s(p,q)-\fpq+3\sign{\pqlf}\right]
\nonumber\\
&&\;\;\;\;\times
\int_{0}^{\infty}Kda_n
\int_{\stackrel{\scriptstyle |\va_j|=\a_j/K}{|\va_n|=a_n}}
\pva
\exp\left[\frac{i\pi K}{2}\left(
\sum_{m=2}^{\infty}\lm+\fpq\va_n^2\right)\right]
\nonumber\\
\shift{
\times\sin\left(\pi\frac{a_n}{q}\right)\onespml,
}
\nonumber
\end{eqnarray}
or, equivalently,
\begin{eqnarray}
\lefteqn{
Z^{({\rm tr})}_{\a_1,\ldots,\a_{n-1}}
(M^{\prime},\cL_{[n]};k)=
}
\label{5.33}\\
&&=
\ztrmk\emsp
\fsq
\exp\frac{i\pi}{2K}\left[12s(p,q)-\fpq+3\sign{\pqlf}\right]
\nonumber\\
&&\;\;\;\;\times
\int\frac{K^2}{4\pi}d^3 \va_n
\int_{|\va_j|=\a_j/K}
\prod_{j=1}^{n-1}\left(\frac{K}{4\pi}
\frac{d^2\va_j}{|\va_j|}\right)
\exp\left[\frac{i\pi K}{2}\left(
\sum_{m=2}^{\infty}\lm+\fpq\va_n^2\right)\right]
\nonumber\\
\shift{
\times\sin\left(\pi\frac{a_n}{q}\right)\onespml,
}
\nonumber
\end{eqnarray}
The integral over $\va_n$ should be calculated in the following way.
We first separate the part of the sum $\sum_{m=2}^{\infty}\lm$ which
is linear in $\va_n$:
\qq
\va_n\cdot\sum_{m=2}^{\infty}\vec{L}_m^{(n)}\adots
\label{5.34}
\qqq
and introduce a new variable $\vx$ instead of $\va_N$:
\qq
\va_N=\vx-\frac{1}{2}\fpql
\va_n\cdot\sum_{m=2}^{\infty}\vec{L}_m^{(n)}\adots
\label{5.35}
\qqq
After substituting eq.~(\ref{5.35}) into the integral~(\ref{5.33}) we
separate the terms of the exponent that do not depend on $\vx$. These
terms form the exponent of the representation~eq.~(\ref{5.1}) for
$Z^{({\rm tr})}_{\a_1,\ldots,\a_{n-1}}
(M^{\prime},\cL_{[n]};k)$. We leave the term $\frac{\pql}{q}\vx^2$ in
the exponent of eq.~(\ref{5.33}) and expand that exponent in all other
terms which are at least quadratic in $\vx$ (there are no linear terms
thanks to the substitution~(\ref{5.35})). This expansion mixes up with
expansions in powers of $\vx$ of two preexponential factors
$1+\spml$ and $\sin\left(\pi\frac{|\va_n|}{q}\right)|\va_n|^{-1}$ (the
latter factor is in fact analytic in $\va_n$ since its expansion
contains only even powers of $|\va_n|$). Thus, similar to the
integral~(\ref{3.*4}), what remains is a bunch of gaussian integrals
over $\vx$. The limit on the powers of $K$ versus powers of $\vx$ in
the expansion of preexponential factor is weaker than that of
eq.~(\ref{3.*4}) (e.g. we now have positive powers of $K$). However it
is easy to see that the main property still holds (cf.
Corollary~\ref{c3.1}):
%
\begin{proposition}
Only a finite number of the polynomials $L_m$ and $P_{m,l}$ of
eq.~(\ref{5.33}) are needed to express a given polynomial $L_m$,
$P_{m,l}$ or a given term in $1/K$ expansion of
$\ztrmpk$ participating in the expression~(\ref{5.1}) for
$Z^{({\rm tr})}_{\a_1,\ldots,\a_{n-1}}
(M^{\prime},\cL_{[n]};k)$.
\label{P5.4}
\end{proposition}
%

To determine what happens to the multivariable Alexander polynomial
under the surgery $\Upq$ on the link component $\cL_n$ we have to find
the contribution of the configuration~(\ref{5.12}) to the
integral~(\ref{5.32}) to the leading order in $K$. In view of
Conjecture~\ref{P5.1}, the integral over $a_n$ is dominated by
the stationary phase point
\qq
a_n^{({\rm st})}=-\frac{q}{\pql}
\sum_{j=1}^{n-1}l_{jn}a_j.
\label{5.36}
\qqq
We need only the 1-loop approximation to this integral:
\begin{eqnarray}
\lefteqn{
Z^{({\rm red})}_{(a_1,\ldots,a_{n-1})}(M^{\prime},\cL_n;k)= }
\label{5.37}\\ &&
\!\!
=\sqrt{\frac{2}{K}}\pi[\ohm]^{-\frac{3}{2}}|\pql|^{-\frac{1}{2}}\,
2i\sin\left(\frac{\sum_{j=1}^{n-1}l_{jn}a_j}
{\pql}\right)\sign{\pql}
\nonumber\\
&&
\!\!
\times
\int_{\stackrel{\scriptstyle |\va_j|=|a_j|}{|\va_n|=|a_n^{({\rm
st})}|}}
\pva\elm\!\!\onespml
\nonumber\\
\shift{
+ \cO(K^{-\frac{3}{2}}),
}
\nonumber
\end{eqnarray}
and we have to take only the contribution of the
configuration~(\ref{5.12}) with $a_n=a_n^{({\rm st})}$ to this
integral. Comparing this expression with eqs.~(\ref{1.14})
and the surgery formula for the 1-loop correction
\qq
e^{iS_1^{({\rm tr})}(M^\prime)}=
|p+\nu q|^{-\frac{3}{2}}\,e^{iS_1^{({\rm tr})}(M)}
\label{3.5}
\qqq
we conclude that
%
\begin{proposition}
If a $\Upq$ surgery on the $n$th component of a link $L$ in \rhs $M$
produces another \rhs $M^{\prime}$, then for the remaining link
$\cL_{[n]}$
\begin{eqnarray}
\lefteqn{
\Delta_A(M^{\prime},\cL_{[n]};
e^{2\pi ia_1},\ldots,e^{2\pi ia_{n-1}})=
2i\sin\left(\frac{\sum_{j=1}^{n-1}l_{jn}a_j}{\pql}\right)
\sign{\pql}
}
\label{5.38}\\
\shift{
\times\Delta_A\left(M,\cL;e^{2\pi ia_1},\ldots,
e^{2\pi ia_{n-1}},\exp\left(-\frac{2\pi iq}{\pql}
\sum_{j=1}^{n-1}l_{jn}a_j\right)\right).
}
\nonumber
\end{eqnarray}
\label{P5.5}
\end{proposition}
%

\nsection{The Link Surgery Formula}
\label{s7}

Now we turn to the subject of our main concern: the surgery formula
for the contribution of the trivial connection to Witten's invariant.
Suppose that a \rhs $M$ contains an $N$-component link $\cL$ and we
perform $U^{(p_j,q_j)}$ surgeries on its components in order to obtain
a new \rhs $M^{\prime}$. Applying the arguments
of Section~3 of \cite{RoI} to eq.~(\ref{5.1}) instead
of eq.~(\ref{2.15}) we conclude that
%
\begin{proposition}
The trivial connection
contribution to Witten's invariants of \rhss $M$
and $M^{\prime}$ connected by $U^{(p_j,q_j)}$ surgeries on components
of a link $\cL$ in $M$, are related by the following equation:
\begin{eqnarray}
\ztrmpk&=&
\ztrmk\exp\left(-\frac{3}{4}i\pi\sign{L^{({\rm tot})}}\right)
\label{5.41}\\
&&\times
\exp\frac{i\pi}{2K}\left[3\sign{L^{({\rm tot})}}+
\sum_{j=1}^{n}\left(12s(p_i,q_i)-\frac{p_i}{q_i}\right)\right]
\nonumber\\
&&\times
\int_{[\va_j=0]}\prod_{j=1}^{n}\left(
\frac{K^2}{4\pi}d^3\va_j \frac{2\sign{q_j}}{\sqrt{2K|q_j|}}
\frac{\sin\left(\pi\frac{|\va_j|}{q_j}\right)}
{|\va_j|}\right)\!\!\onespml
\nonumber\\
&&\hspace*{5cm}\times
\exp\left(\frac{i\pi K}{2}\sum_{i,j=1}^{n}L^{({\rm tot})}_{ij}
\va_i\cdot\va_j\right).
\nonumber
\end{eqnarray}
The symbol $\int_{[\va_j=0]}$ means that we take only the contribution
of the stationary phase point $\va_j=0,\;\;1\leq j\leq n$, which
should be calculated in the following way: all the factors except for
the last exponential should be expanded in powers of $\va_j$ and then
the gaussian integrals with polynomial prefactors should be calculated
one by one.
\label{P5.6}
\end{proposition}
%
Although in contrast to eq.~(\ref{3.*4}) there will be positive powers
of $K$ in the preexponential series, still
%
\begin{corollary}
Only a finite number of
the polynomials $L_m$ and $P_{m,l}$ are required
to express $\ztrmpk$ at a given order in $1/K$ expansion.
\label{C5.1}
\end{corollary}
%

We can use eq.~(\ref{5.41}) in order to derive explicit surgery
formulas for the first two loop corrections to $\ztrmk$:
\qq
e^{iS_1^{({\rm tr})}(M^{\prime})}=
\left|\det L^{({\rm tot})}\prod_{j=1}^{n}q_n\right|^{-\frac{3}{2}}
e^{iS_1^{({\rm tr})}(M)},
\label{5.42}
\qqq
which is consistent with eq.~(\ref{3.7}) and
\qq
S_2^{({\rm tr})}(M^{\prime})&=&
S_2^{({\rm tr})}(M)+3\Delta_{\rm CW},
\label{5.43}
\qqq
here
\begin{eqnarray}
\Delta_{\rm CW}&=&\frac{1}{4}\sign{L^{({\rm tot})}}
+\sum_{j=1}^{n}\left((s(p_j,q_j)-
\frac{1}{12}\left(\frac{p_j}{q_j}+l_{jj}\right)\right)
\label{5.44}\\
&&-\frac{1}{\pi^2\det L^{({\rm tot})}}
\left[-\frac{1}{2}\sum_{i,j=1}^{n}\left(p_{ij}
-\frac{1}{6}\frac{\pi^2}{q_i^2}\delta_{ij}\right)\partial_{l_{ij}}
+\frac{3}{2}\sum_{i_1,i_2,j_1,j_2=1}^{n}
\lf_{i_1 i_2,j_1 j_2}
\partial_{l_{i_1 j_1}}
\partial_{l_{i_2 j_2}}
\right.
\nonumber\\
\sshift{
\left.
+\frac{3}{4}\sum_{i_1,i_2,i_3,j_1,j_2,j_3=1}^{n}
\lt_{i_1 i_2 i_3}\lt_{j_1 j_2 j_3}
\partial_{l_{i_1 j_1}}\partial_{l_{i_2 j_2}}
\partial_{l_{i_3 j_3}}\right]\det L^{({\rm tot})}.
}
\nonumber
\end{eqnarray}
In view of eq.~(\ref{3.8}) we assume that
\qq
\lambda_{\rm CW}(M^{\prime})=\lambda_{\rm CW}(M)+
\Delta_{\rm CW}.
\label{5.45}
\qqq
We did not compare eqs.~(\ref{5.45}), (\ref{5.44}), (\ref{5.25})
and~(\ref{5.27}) directly to the surgery formula of \cite{Le2}.
However the latter formula was derived from Walker's surgery formula
by using the properties~(\ref{5.31}) and~(\ref{5.38}) of the
multivariable Alexander polynomial. Since our formula also satisfies
these properties, we assume that it is consistent with the results of
\cite{Le2}.

\nsection{Discussion}
The results of this paper are based on the Reshetikhin's
formula~(\ref{5.1}) which separates the exponent of order $K$ from
the preexponential factor of order at most $K^0$. This separation
allowed us to extract the large $k$ asymptotics of the Jones
polynomial of a link and of the link surgery formula~(\ref{1.12}).

Assuming that Conjecture~\ref{cf3.1} is correct we see the relation
between the leading part of the multivariable Alexander polynomial
when its arguments are close to 1, and Milnor's linking numbers of
the knot. Slighly generalizing the results of~\cite{Kl} and
{}~\cite{FrKl} we may say that the Alexander
polynomial and Milnor's linking numbers are the algebraic tools for
the study of irreducible deformations of reducible flat connections
in the knot complement: the zeros of the Alexander polynomial
indicate the points where the deformation can be carried out and
Milnor's numbers determine the possible directions of the
deformation.

The surgery formula for the loop corrections $S_n^{({\rm tr})}$ to
the trivial connection contribution to Witten's invariant of a
rational homology sphere as defined by eq.~(\ref{1.13}) was derived
in~\cite{RoI} at the ``physical'' level of rigor. The extension of
this formula to links provided by the Proposition~\ref{P5.6} gives it
a better chance to acquire a rigorous mathematical proof. In other
words, the invariance of the r.h.s. of eq.~(\ref{5.41}) under Kirby
moves has to be established.

Eq.~(\ref{5.41}) is a surgery formula for the perturbative invariants
$S_n^{({\rm tr})}$ defined in canonical framing. The same invariants
can be calculated through Feynman diagrams which require a certain
regularization~\cite{Ta}. The relation between this regularization
and the choice of framing still remains to be understood.

Another open question is a calculation of the contributions of
nontrivial connections as well as the extension of this discussion
beyond the rational homology spheres. Some experimental results on
Witten's invariant for these cases are provided in~\cite{FrGo},
{}~\cite{Je}, ~\cite{Ro1} and ~\cite{RoI}, while a study of Casson's
invariant of the manifolds with nontrivial rational homology was
carried out in~\cite{Le2}. However, all these results seem to require
a more detailed analysis.

\section*{Acknowledgements}

I am thankful to D.~Auckley, D.~Bar-Natan, D.~Freed, C.~Lescop,
E.~Klassen, and K.~Walker
for discussing the various subjects of this work. I am
especially thankful to N.~Reshetikhin, O.~Viro and A.~Vaintrob for
many consultations and encouragement.

This work was supported by the National Science Foundation
under Grant No. PHY-92 09978.

\nappendix{1}
\label{A*2}

We are going to derive Reshetikhin's formula~(\ref{5.1}) for the type
$(n,mn)$ torus link $\tl$. This is a very simple link which consists
of $n$ parallel components which are twisted $m$ times. Its Jones
polynomial is easy to calculate:
\begin{eqnarray}
\zatl&=&-\frac{i^{\frac{1}{2}}}{K\sqrt{m}}
\exp\left[-\frac{i\pi}{K}\frac{m}{2}\left(
1+\sum_{j=1}^{n}(\a_i^2-1)\right)\right]
\label{A2.1}\\
&&\times
\sum_{\b=1}^{K-1}\frac{\prod_{j=1}^{n}
\sin\left(\frac{\pi}{K}\a_i\b\right)}
{\sin^{n-1}\left(\frac{\pi}{K}\b\right)}
\sum_{l=0}^{m-1}\sum_{\mu=\pm 1}\mu
\exp\left[-\frac{i\pi}{2Km}(\b+2Kl+\mu)^2\right].
\nonumber
\end{eqnarray}
The sum over $l$ is a nuisance because it is nowhere present in the
r.h.s. of eq.~(\ref{5.1}). However by applying the methods of
Section~4 of~\cite{RoI}
we can show that the contribution of the terms  with
$l\neq 0$ is related only to irreducible connections which appear only
when the values of the phases $\a_i/K$ are large enough (the $n+1$
numbers $\a_i/K$ and $2l/m$ should satisfy ``polygon inequality''
conditions). Ultimately for small values of $\a_i/K$ we can use the
expression
\begin{eqnarray}
\zatl&=&\frac{2i^{\frac{3}{2}}}{K\sqrt{m}}
\exp\left[-\frac{i\pi}{K}\frac{m}{2}\left(1-n+\frac{1}{m^2}+
\sum_{j=1}^{n}\a_j\right)\right]
\label{A2.2}\\
&&\times
\int_{\stackrel{\scriptstyle -\infty}
{[0\leq\beta\leq K]}}^{+\infty}
\frac{\prod_{j=1}^{n}
\sin\left(\frac{\pi}{K}\a_i\b\right)}
{\sin^{n-1}\left(\frac{\pi}{K}\b\right)}
\sin\left(\frac{\pi}{K}\frac{\beta}{m}\right)
\exp\left(-\frac{i\pi}{2K}\frac{\beta^2}{m}\right).
\nonumber
\end{eqnarray}
A simple formula
\qq
\int_{|\va|=\frac{\a}{K}}\frac{K}{4\pi}
\frac{d^2\va}{|\va|}
\exp\left(i\pi K\va\cdot\vb\right)=
\frac{\sin(\pi\a|\vb|)}{\pi|\vb|}
\label{A2.3}
\qqq
allows us to rewrite eq.~(\ref{A2.2}) as
\begin{eqnarray}
\zatl&=&Z(S^3;k)\left(\frac{i}{2}\frac{K}{m}\right)^{\frac{3}{2}}
\exp\left[\frac{i\pi}{2K}m\left((n-1)-m^{-2}\right)\right]
\label{A2.4}\\
&&\times\spint\int d^3\vb
\left(\frac{\pi|\vb|}{\sin(\pi|\vb|)}\right)^{n-1}
\frac{\sin\left(\pi\frac{|\vb|}{m}\right)}
{\pi\frac{|\vb|}{m}}
\nonumber\\
&&
\times\exp\left[-\frac{i\pi K}{2}
\left(m\sum_{j=1}^{n}\va_j^2-2\vb\cdot\sum_{j=1}^{n}\va_j
+\frac{\vb^2}{m}\right)\right]
\nonumber
\end{eqnarray}
After changing the integration variable from $\vb$ to
\qq
\vx=\vb-m\sum_{j=1}^{n}\va_j,
\label{A2.5}
\qqq
expanding
the preexponential factor in powers of $\vx^2$ and calculating
gaussian integrals over $\vx$ we obtain the formula
\begin{eqnarray}
\lefteqn{
\zatl=
}\label{A2.6}\\
&&
=Z(S^3;k)\spint
\exp\left[\frac{i\pi K}{2}m
\sum_{\stackrel{\scriptstyle i,j=1}{i\neq j}}^{n}
\va_i\cdot\va_j\right]
\,\exp\left[\frac{i\pi}{2K}m(n-1-m^{-2})\right]
\nonumber\\
\shift{
\times
\left.
\frac{\left(\frac{\pi}{K}\right)}
{\sin\left(\frac{\pi}{K}\right)}
\sum_{l=0}^{\infty}\left(-\frac{im}{2\pi K}\right)^l
\frac{(2l+1)!}{l!\,(2l)!}
\;\partial_y^{(2l)}
\left[\left(\frac{\pi y}{\sin(\pi y)}\right)^{n-1}
\frac{\sin\left(\pi\frac{y}{m}\right)}
{\left(\pi\frac{y}{m}\right)}\right]
\right|_{y=m\left|\sum_{j=1}^{n}\va_j\right|}.
}
\nonumber
\end{eqnarray}
This is Reshetikhin's formula~(\ref{5.1}). In particular,
\qq
l_{ij}=m\;\;\;\;{\rm for}\;\;i\neq j,\;\;\;\;l_{jj}=0
\label{A2.7}
\qqq
and
\qq
\lm=0\;\;\;\;{\rm for}\;\;m\geq 3,
\label{A2.8}
\qqq
(this seems to be a general property of torus links), while
\begin{eqnarray}
\lefteqn{
1+\spml=\exp\left[\frac{i\pi}{2K}m(n-1-m^{-2})\right]
\frac{\left(\frac{\pi}{K}\right)}
{\sin\left(\frac{\pi}{K}\right)}
}
\label{A2.9}\\
\shift{
\times
\left.
\sum_{l=0}^{\infty}\left(-\frac{im}{2\pi K}\right)^l
\frac{(2l+1)!}{l!\,(2l)!}
\;\partial_y^{(2l)}
\left[\left(\frac{\pi y}{\sin(\pi y)}\right)^{n-1}
\frac{\sin\left(\pi\frac{y}{m}\right)}
{\left(\pi\frac{y}{m}\right)}\right]
\right|_{y=m\left|\sum_{j=1}^{n}\va_j\right|}.
}
\nonumber
\end{eqnarray}

It is easy to check the relation~(\ref{5.22}) between eq.~(\ref{A2.6})
and the Alexander polynomial if we recall that
\qq
\Delta_A(S^3,\tl;e^{2\pi ia_1},\ldots,e^{2\pi ia_n})=
(2i)^{n-2}\frac{\sin^{n-1}\left(\pi m\sum_{j=1}^{n}a_j\right)}
{\sin\left(\pi\sum_{j=1}^{n}a_j\right)}
\label{A2.10}
\qqq
and that in our case
\qq
-i\emmind|\det M^{\prime\prime}|^{\frac{1}{2}}=
i^{n-2}m^{n-1}\left(\sum_{j=1}^{n}a_j\right)^{2-n}
\prod_{j=1}^{n}a_j.
\label{A2.11}
\qqq

The torus link $\tl$ provides an example of existence of irreducible
flat connections in the link complement even for arbitrarily small
phases $|\va_j|$. Eq.~(\ref{5.11}), which in view of eq.~(\ref{A2.8})
is exactly valid for small phases, is reduced to a condition
\qq
\left(\sum_{i=1}^{n}\va_i\right)\times\va_j=0,\;\;\;\;\;1\leq j\leq n,
\label{A2.12}
\qqq
which is obviously satisfied if
\qq
\sum_{j=1}^{n}\va_j=0.
\label{A2.13}
\qqq
The necessary and sufficient condition for the existence of this
configuration is that the phases $|\va_j|$ satisfy ``polygon
inequalities'':
\qq
\sum_{\stackrel{\scriptstyle i=1}{i\neq j}}^{n}
|\va_i|\leq|\va_j|,\;\;\;\;\;1\leq j\leq
n.
\label{A2.14}
\qqq
These inequalities can indeed be satisfied even for arbitrarily small
phases. Note that the extremal cases of these inequalities, i.e.
\qq
|\va_j|=\sum_{\stackrel{\scriptstyle i=1}{i\neq j}}^{n}
|\va_i|,
\label{A2.15}
\qqq
are parallel configurations and also zeros of the
multivariable Alexander polynomial~(\ref{A2.10}).

It is possible to combine
the calculations of Appendix in~\cite{RoI}
and eq.~(\ref{A2.6}) into
Reshetikhin's representation of the Jones polynomial of a general
$p$-component torus link $\cL_{(mp,np)}$ ($m$ and $n$ are coprime). We
present here the result without derivation (it is similar to the one
for eq.~(\ref{A2.6})):
\begin{eqnarray}
\lefteqn{
Z_{\a_1,\ldots,\a_p}(S^3,\cL_{(mp,np)};k)=
}\label{A2.16}\\
&&
=Z(S^3;k)
\int_{|\va_j|=\frac{\a_j}{K}}
\prod_{j=1}^{p}
\left(\frac{K}{4\pi}\frac{d^2\va_j}{|\va_j|}\right)
\exp\left[\frac{i\pi K}{2}mn
\sum_{\stackrel{\scriptstyle i,j=1}{i\neq j}}^{p}
\va_i\cdot\va_j\right]
\,\exp\left[\frac{i\pi}{2K}
\frac{m^2 n^2 p-m^2-n^2}{mn}
\right]
\nonumber\\
\shift{
\times
\left.
\frac{\left(\frac{\pi}{K}\right)}
{\sin\left(\frac{\pi}{K}\right)}
\sum_{l=0}^{\infty}
\left(2\pi iK\right)^{-l}
\frac{(2l+1)!}{l!\,(2l)!}
\;\partial_y^{(2l)}
\left[
\left(\frac{\pi mny}{\sin(\pi mny)}\right)^{p}
\frac{\sin(\pi my)}{\pi my}\;
\frac{\sin(\pi ny)}{\pi ny}
\right]
\right|_{y=\left|\sum_{j=1}^{p}\va_j\right|}.
}
\nonumber
\end{eqnarray}
Eqs.~(A1.4) of~\cite{RoI}
and~(\ref{A2.6}) are particular cases of this
equation (set p=1 or set n=1 and put n instead of p).

The Alexander polynomial of $\cL_{(mp,np)}$ is
\qq
\Delta_A(S^3,\cL_{(mp,np)}
;e^{2\pi ia_1},\ldots,e^{2\pi ia_p})=
(2i)^{p-2}
\frac{\sin^p\left(\pi mn\sum_{j=1}^{p}a_j\right)}
{\sin\left(\pi m\sum_{j=1}^{p}a_j\right)
\sin\left(\pi n\sum_{j=1}^{p}a_j\right),}
\label{A2.17}
\qqq
its relation to eq.~(\ref{A2.16}) is easy to observe.

\nappendix{2}
\label{A*3}

Here we will briefly review the structure of flat connections in a
link complement and show that eq.~(\ref{5.11}) describes them
approximately in close vicinity of the trivial connection. For more
details on the structure of flat connections in the link complement
and their relation to Milnor's linking numbers see for
example~\cite{Ro2}.

Consider an
$n$-component link $\cL$ in $S^3$.
We use Wirtinger's presentation for the group $\gpi$.
%
%
We project the link $\cL$ onto a
2-dimensional plane and denote as $L_{i,j}$ the pieces into which a
link component is split
when it is overcrossed. With each such piece we
associate an element $g_{i,j}\in\gpi$. These elements generate the
whole group $\gpi$ modulo certain relations. Let
$P_{i,j}^{k,l}$
be a
crossing point where a piece $L_{k,l}$ overcrosses a junction of two
pieces $L_{i,j}$ and $L_{i,j+1}$. Let
$\sign{P_{i,j}^{k,l}}$
be a signature of this crossing. In other words,
$\sign{P_{i,j}^{k,l}}$
is either $+1$ or $-1$ depending on mutual orientation of $L_i$ and
$L_k$ at the point of crossing. The linking number of two link
components can be expressed in terms of the signatures of crossings:
\qq
l_{ik}=\sum_{j,l}\sign{P_{i,j}^{k,l}}.
\label{A3.0}
\qqq
The relation between the group elements corresponding to the crossing
point $P_{i,j}^{k,l}$ is
\qq
g_{i,j+1}=g_{k,l}^{\sign{P_{i,j}^{k,l}}}
g_{i,j}g_{k,l}^{-\sign{P_{i,j}^{k,l}}}.
\label{A3.1}
\qqq
The relations~(\ref{A3.1}) describe the structure of $\gpi$.

Suppose that we have a one-parametric family of homomorphisms
\begin{eqnarray}
&\gpi\rightarrow G,
\nonumber\\
&
g_{i,j}\mapsto\exp\left(
\sum_{m=1}^{\infty}\l_{i,j}^{(m)}t^{m}\right),\;\;\;
t\geq 0,
\label{A3.2}
\end{eqnarray}
here $G$ is a Lie group and $\l_{i,j}^{(m)}$ are elements of its Lie
algebra. The homomorphisms~(\ref{A3.2}) describe (up to a
conjugation) a family of flat connections in $\smin$
which includes the trivial connection at $t=0$.

We substitute the images of the homomorphism~(\ref{A3.2}) into the
relations~(\ref{A3.1}) and expand them in powers of $t$. At order
$t^1$ we observe that the elements $\l_{i,j}^{(1)}$ do not depend on
$j$, so we denote them simply as
\qq
\l_i=\l^{(1)}_{i,j}.
\label{A3.3}
\qqq
At order $t^2$ we get a relation
\qq
\l_{i,j+1}^{(2)}-\l_{i,j}^{(2)}=\sign{P_{i,j}^{k,l}}
[\l_k,\l_i].
\label{A3.4}
\qqq
If we go around a link component
$\cL_i$ and add together all the
relations~(\ref{A3.4}), then we arrive at the equation
\qq
\sum_{k,l,j}\sign{P_{i,j}^{k,l}}
[\l_k,\l_i]=0,
\label{A3.5}
\qqq
which in view of eq.~(\ref{A3.0}) is equivalent to eq.~(\ref{5.11})
for the case of $G=SU(2)$.
%
\begin{proposition}
The stationary points of the phase in Reshetikhin's
formula~(\ref{5.1}) are in one-to-one correspondence with the flat
connections on the link component in the linear approximation around
the trivial connection.
\label{pA3.1}
\end{proposition}

\end{document}